\documentclass[fleqn,12pt,twoside]{article}
\usepackage{espcrc1}
\usepackage{epsfig}

\usepackage{graphicx}
\usepackage[figuresright]{rotating}

\newcommand{\AmS}{{\protect\the\textfont2
  A\kern-.1667em\lower.5ex\hbox{M}\kern-.125emS}}

\title{Thermal description of transverse-momentum spectra at RHIC
       \thanks{Supported in part by the Polish State Committee for
  Scientific Research, grant 2 P03B 09419.}}

\author{Wojciech Florkowski
      and Wojciech Broniowski   \\   \vspace{0.3cm}
H. Niewodnicza\'nski Institute of Nuclear Physics, 
        \\  ul. Radzikowskiego 152, 31-342 Krak\'ow, Poland }

\begin{document}

\maketitle

\begin{abstract}
We show that the transverse-momentum spectra of all hadrons measured
at RHIC, including the hyperons, are described very well in a thermal
model assuming the simultaneous chemical and thermal freeze-outs. The
model calculation takes into account all hadronic resonances and uses
a simple parametrization of the freeze-out hypersurface.
\end{abstract}

\vspace{0.5cm}

We present a simple model describing the $p_\perp$-spectra of hadrons
measured at RHIC at $\sqrt{s_{NN}}$ = 130 GeV. Our approach is a
combination of the thermal model, used frequently in the studies of
the relative hadron yields
\cite{raf,cest,pbmsps,yg,becatt,gaz,raf0,pbmrhic,fbm}, with a model of
the hydrodynamic expansion of matter at freeze-out. The main
assumptions of the model \cite{wbwf,str,hirsch} are as follows: {\it
i)} the chemical freeze-out and the thermal freeze-out occur
simultaneously, which means that we neglect elastic rescattering after
the chemical freeze-out, {\it ii)} all hadronic resonances are
included in both the calculation of the hadron multiplicities and the
spectra, and {\it iii)} a simple form of the freeze-out hypersurface
is proposed, which is a generalization of the Bjorken model
\cite{bjorken} (see also
\cite{baym,nikolaev,siemens,schnedermann,BL,rischke}),
\begin{equation}
\tau = \sqrt{t^2-x^2-y^2-z^2} = \hbox{const}.
\label{tau}
\end{equation}
The hydrodynamic flow on the freeze-out hypersurface (\ref{tau}) is
taken in the form resembling the Hubble law,
\begin{equation}
u^\mu = {x^\mu \over \tau} = {t \over \tau} \left(1, {x \over t},
{y \over t},{z \over t}\right).
\label{umu}
\end{equation}

Recently, new arguments have been accumulated in favor of our first
assumption. The measurements of the $K^*(892)$ states by the STAR
Collaboration \cite{starKstar} indicate that either the daughter
particles from the decay $K^*(892) \rightarrow K \pi $ do not
rescatter or the expansion time between the chemical and thermal
freeze-out is short (smaller than the $K^*(892)$ lifetime, $\tau$=4
fm/c). Moreover, the measured yield of $K^*(892)$ fits very well to
the pattern obtained from the thermal analysis of the ratios of hadron
abundances. This fact suggests again a short
expansion time between the two freeze-outs.  The assumption about the
single freeze-out also solves the antibaryon puzzle \cite{rapp}. Since
the annihilation cross section for $p {\bar p}$ pairs is much larger
than the elastic cross section, most of the protons would annihilate
with antiprotons during the long way from the chemical to the thermal
freeze-out. Such effect is not seen. In addition, let us mention that
the single-freeze-out scenario is natural if the hadronization process
occurs in such a way that neither elastic or inelastic processes are
effective. An example here is the {\it sudden-hadronization} model of
Ref. \cite{raf1}.
\begin{figure}[t]
\begin{center}
\epsfysize=12.50cm \mbox{\epsfbox{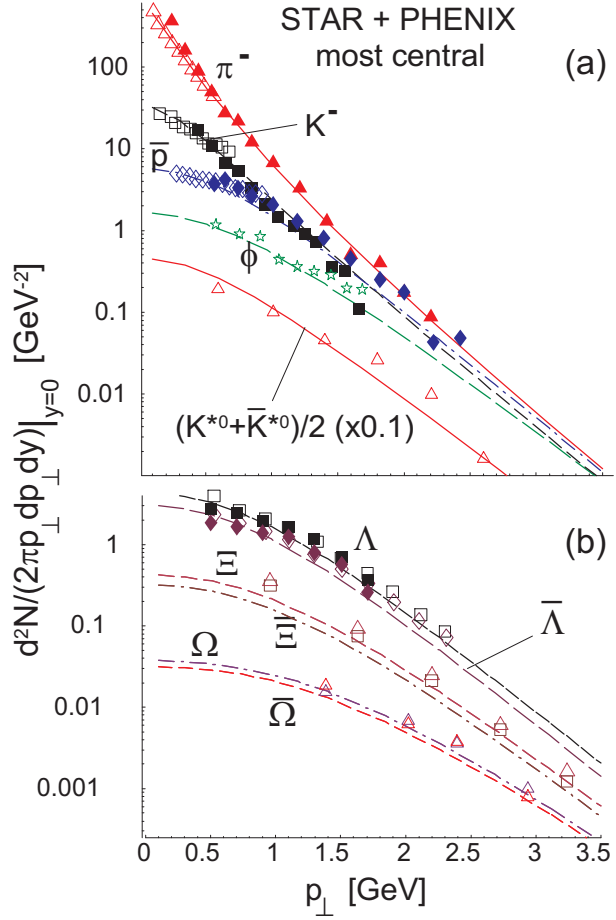}}
\end{center}
\label{minb}
\vspace{-1.5cm}
\caption{ The $p_\perp$-spectra at midrapidity of
$\pi^-, K^-, {\bar p}\,, \phi$ and $K^*(892)$ in part (a), and of the
hyperons $ \Lambda\,, \Xi\,$ and ${\Omega}$ in part (b).  The model
calculation is compared to the PHENIX (filled symbols) and STAR (open
symbols) most central data
\cite{starKstar,phenix,harris,starpbar,starphi,starLambda,phenixLambda,starXi,starOmega}
from Au + Au collisions at $\sqrt{s_{NN}}$ = 130 GeV. Both the data
and the theoretical curves are absolutely normalized (they include
full feeding from the weak decays).  }
\label{fig1}
\end{figure}

Our model has two thermodynamic and two geometric (expansion)
parameters.  The two thermodynamic parameters, $T$ = 165 MeV and
$\mu_B$ = 41 MeV, were obtained from the analysis of the ratios of the
hadron multiplicities measured at RHIC \cite{fbm}. In this calculation
the grand-canonical ensemble was used without the strangeness
suppression factor ($\gamma_s$=1). Since the particle ratios depend
weakly on the centrality of the collision, we treat the thermodynamic
parameters as the universal parameters (independent of centrality).
The two geometric parameters are $\tau$ of Eq. (\ref{tau}) and
$\rho_{\rm max}$.  The parameter $\rho_{\rm max}$ determines the
transverse size of the firecylinder at the freeze-out,
\begin{equation}
\rho = \sqrt{x^2 + y^2} \le \rho_{\rm max}.
\label{rhomax}
\end{equation}
In the natural way, the values of $\tau$ and $\rho_{\rm max}$ depend
on the considered centrality class of events. For the minimum-bias
data, which average over centralities, we find: $\tau = 5.55 \hbox{
fm}$ and $\rho_{\max} = 4.50 \hbox{ fm}$, whereas for the most central
collisions we find: $\tau = 7.66 \hbox{ fm}$ and $\rho_{\max} = 6.69
\hbox{ fm}$ \cite{wbwf}. The calculation of the spectra (and
determination of the geometric parameters) is based on the standard
Cooper-Frye formalism. The details of our method, especially of the
technical problems concerning the treatment of the resonances, are
given in the Appendix of Ref. \cite{str}.

In Fig. 1 we show our results for the most central collisions. In the
upper part (a) we show the spectra of pions, kaons, antiprotons, the
$\phi$ mesons, and the $K^*(892)$ mesons. In the lower part (b) we
show the spectra of the hyperons $\Lambda, \Xi$ and $\Omega$. The
model calculation agrees very well with the data. Note, e.g., the
convex shape of the pion spectrum, crossing of the pion and the
antiproton spectra at $p_\perp \sim$ 2 GeV, and the good reproducing
of the $\Omega$ spectrum.  The good agreement between the model
calculation and the data supports strongly the idea of thermalization
of the hadronic matter produced at RHIC.  Let us emphasize that the
expansion parameters were fitted in Ref. \cite{wbwf} to the spectra of
pions, kaons and protons only. The spectra of other particles were
calculated with the same values of the parameters, hence, they are
predictions of our model. In view of this fact, the good agreement of
the $\Omega$ spectrum, predicted before the data were available, is
highly non-trivial, especially in the context of the SPS results
\cite{hirsch}.
\footnote{The newest data on $\Xi$ production shown by Castillo during
this conference have smaller overall normalization and also agree very
well with our model calculation.}

A characteristic feature of our approach is a rather high decoupling
temperature $T \sim$ 165 MeV. However, the (inverse) slope parameters
corresponding to this temperature are lowered by the decays of the
resonances \cite{fbm}. This ``cooling'' of the spectrum by the decays of
the resonances explains the difference between the high temperature of
the chemical freeze-out and a smaller ``apparent'' temperature
inferred from the shape of the spectra. We note that a similar high
decoupling temperature has been found in the full hydrodynamic
calculation of Ref. \cite{finland}, where also a complete set of
hadronic resonances is employed. It remains a challenge to check
whether our particular freeze-out conditions (shape of the freeze-out
hypersurface and Hubble flow) may be obtained as the final stage of
hydrodynamic evolution.  First steps in this direction have been
already made \cite{csorgo}.

Let us make a few comments about the size of our geometric parameters.
Translated to the measured HBT radii, $R_{\rm out}$ and $R_{\rm
side}$, they turn out to be too small. This problem can be
circumvented by the inclusion of the excluded-volume corrections
\cite{yg} which affect only the overall normalization of the
spectra. If we rescale $\tau$ and $\rho_{\rm max}$ by about 30\%, we
obtain a satisfactory agreement with the HBT data. On the other hand,
the ratio $R_{\rm out}/R_{\rm side}$ is close to unity in our model,
independently of the excluded-volume corrections. The approximate
equality of these two radii follows
in our model from the fact that the time extension of our system at
freeze-out is much shorter than its space extension.

In conclusion, we want to stress that a simple thermal model (with
altogether four parameters) reproduces the transverse-momentum spectra
of all hadrons which have been measured so far at RHIC. This fact
brings strong evidence for thermalization of hadronic matter at RHIC
and, possibly, indicates that a thermalized system of quarks and
gluons was formed at the earlier stages of the collisions.

\vspace{0.25cm}

We thank Boris Hippolyte for pointing out his talk \cite{starOmega}
with the experimental results.

\end{document}